\documentclass[conference]{IEEEtran}
\IEEEoverridecommandlockouts
\usepackage{algorithm}
\usepackage{algpseudocode}
\usepackage{amsmath}
\usepackage{amsfonts}
\usepackage{graphicx}
\usepackage{epsfig}
\usepackage{cite}
\usepackage{txfonts}
\usepackage{relsize}
\usepackage{color}
\usepackage{color}
\usepackage{CJK}
\usepackage{times}
\usepackage{multicol}
\usepackage{stfloats}
\usepackage{float}
\usepackage{lipsum}
\usepackage{bm}
\usepackage{graphics}

\ifCLASSINFOpdf
\else
\fi

\begin{document}

\title{Two-Phase Channel Estimation for \\RIS-Aided Cell-Free Massive MIMO with Electromagnetic Interference
\thanks{This work was supported by the Research Grants Council under the Area of Excellence scheme grant AoE/E-601/22-R.}}

\author{Jun Qian, Chi Zhang, Khaled B. Letaief, and Ross Murch\\
Dept. of ECE, The Hong Kong University of Science and Technology, Kowloon, Hong Kong
\\Email: eejunqian@ust.hk, czhangcc@connect.ust.hk, eekhaled@ust.hk, eermurch@ust.hk}

\maketitle
{\begin{abstract}

This work considers a reconfigurable intelligent surface (RIS)-aided cell-free massive multiple-input multiple-output (MIMO) system with RIS spatial correlation and electromagnetic interference (EMI). We propose a two-phase channel estimation scheme with fractional power control-aided pilot assignment to improve the estimation accuracy and system performance of RIS-aided cell-free massive MIMO systems. Additionally, we derive the closed-form expressions of the downlink spectral efficiency (SE) with conjugate beamforming to evaluate the impact of EMI among RIS elements on the system performance. Numerical results validate that the proposed two-phase scheme can compensate for the performance degradation caused by EMI in terms of estimation accuracy and downlink SE. Moreover, the benefits of introducing RISs and increasing access points (APs) are illustrated. 

\end{abstract}

\begin{IEEEkeywords}
Cell-free massive MIMO, electromagnetic interference, reconfigurable intelligent surface, spatial correlation, spectral efficiency.
\end{IEEEkeywords}}

\maketitle

\section{Introduction}

The past few decades have witnessed an exponential
growth in wireless communication demand \cite{7827017,9665300}. To satisfy the transmission requirements for future sixth-generation wireless
(6G) communications, several new technologies are developed\cite{10167480,9665300,10225319}. Among them,
cell-free massive multiple-input multiple-output (MIMO) and reconfigurable intelligent surface (RIS) have attracted great attention \cite{9665300,10167480}. As a disruptive technology, cell-free massive MIMO has the potential to enhance
the quality-of-service (QoS) of cell-edge users and mitigate inter-cell interference\cite{10225319,9737367}.
Meanwhile, RISs can improve the transmission performance by shaping the electromagnetic-level radio waves without active power amplifiers or digital signal processing methods \cite{9665300,10167480}. 

Cell-free massive MIMO and RIS are complementary  
and can be integrated to improve the system performance \cite{10167480,9875036}. As a consequence, recent research has focused on exploiting the advantages of the integration of these two technologies \cite{9665300,10167480,9322151,9875036,10225319}. For instance, \cite{9665300} studied the RIS phase shift design optimization to minimize channel estimation
error in RIS-aided cell-free massive MIMO systems.  \cite{10225319} investigated the hardware-impaired
RIS-aided cell-free massive MIMO systems while introducing a modified ON/OFF channel estimation scheme to enhance the system performance. Moreover, \cite{9665300,10264149} showed the importance of investigating the RIS spatial correlation between closely spaced adjacent elements.
The authors in \cite{9598875} pointed out the significant effect of electromagnetic interference (EMI) on the RIS-aided system performance, and this motivated authors in
\cite{10167480} to study the uplink performance of RIS-aided
cell-free massive MIMO systems experiencing EMI.
In practice, EMI and RIS spatial correlation can be non-negligible in RIS-aided cell-free massive MIMO systems; thus, the corresponding performance limits should be carefully investigated \cite{10133717,10167480}.

This work studies the system performance of spatially correlated RIS-aided cell-free massive MIMO, considering EMI. Specifically, we introduce a two-phase channel estimation scheme with fractional power control-aided pilot assignment to introduce channel estimation accuracy gain and compensate for the performance degradation. Moreover, the closed-form downlink spectral efficiency (SE) expressions by applying conjugate beamforming in the downlink transmission are derived. Our numerical and analytical results indicate the benefits of the novel two-phase channel estimation scheme in terms of channel estimation accuracy and system performance.

{\textit{Organization:}} Section II describes the spatially correlated RIS-aided Cell-free massive MIMO system model with EMI. Section III introduces a two-phase channel estimation scheme. Section IV derives the closed-form downlink SE expressions. Section V verifies the numerical results, and Section VI summarizes the work. 

\section{System Model}
\subsection{Channel Model}
This paper considers a RIS-aided cell-free massive MIMO system under time-division duplex (TDD) operation, where $M$ $N$-antenna access points (APs) serve $K$ single-antenna users simultaneously. All APs are connected to a central processing unit (CPU) through ideal fronthaul links\cite{9940169,10571171}. The communications between APs and users are enhanced by $J$ $L$-element RISs\cite{10225319} with $L=L_v\times L_h$ in which $L_v$ and $L_h$ are the respective numbers of row and column elements per RIS.

For analytical tractability, we assume that the associated channels experience Rayleigh fading \cite{9875036,10225319}. The aggregate uplink channel from the $k$-th user to the $m$-th AP is formulated as
\begin{equation}
     \displaystyle \textbf{g}_{mk}=\textbf{g}_{mk}^{\text{d}}+\sum\limits_{j=1}^{J}\textbf{g}_{mkj}^{\text{c}},
     \label{aggregate_uplink_channel}
   \end{equation}
where $\textbf{g}_{mk}^{\text{d}}\in \mathbb{C}^{N\times1}$ represents the direct channel from the $k$-th user to the $m$-th AP experiencing uncorrelated Rayleigh fading\cite {9875036}, given by
\begin{equation}
     \displaystyle \textbf{g}_{mk}^{\text{d}}=\sqrt{\beta_{mk}^{\text{d}}}\textbf{v}_{mk}^{\text{d}},
     \label{direct_uplink_channel}
   \end{equation}
where $\beta_{mk}^{\text{d}}$ is the large-scale fading coefficient, $\textbf{v}_{mk}^{\text{d}}\sim \mathcal{CN}(\textbf{0},\textbf{I}_{N})$ represents the independent fast-fading channel.

The cascaded channel between the $k$-th user and the $m$-th AP via the $j$-th RIS is introduced by $\textbf{g}_{mkj}^{\text{c}}$, written as
\begin{equation}
     \displaystyle
\textbf{g}_{mkj}^{\text{c}}=\textbf{g}_{mj}^{\text{c}}\Theta_{j}\textbf{g}_{kj}^{\text{c}},
    \label{cascaded_uplink_channel_via_RIS}
   \end{equation}
where $\Theta_j=\text{diag}(\alpha_{j,1},\alpha_{j,2},...,\alpha_{j,L})\in\mathbb{C}^{L\times L}$ shows the phase shift matrix of the $j$-th RIS. $\alpha_{j,l}=u_{j,l}e^{i\theta_{jl}}$ denotes the reflection coefficient of the $l$-th element at the $j$-th RIS, with the amplitude reflection coefficient $u_{j,l}\in[0,1]$ and the induced phase shift $\theta_{jl}\in[0,2\pi]$ \cite{10167480,10225319}. Since our proposed model capturing the RIS spatial correlation \cite{10167480,9665300}, the channel from the $j$-th RIS to the $m$-th AP, $\textbf{g}_{mj}^{\text{c}}\in \mathbb{C}^{N\times L}$, is given by 
\begin{equation}
     \displaystyle \textbf{g}_{mj}^{\text{c}}=\sqrt{{\beta_{mj}^{\text{c}}}}\textbf{v}_{mj}^{\text{c}}\textbf{R}_{mj}^{1/2},
     \label{channel_RIS_AP}
   \end{equation}
   where $\beta_{mj}^{\text{c}}$ is the large-scale fading coefficient, $\textbf{v}_{mj}^{\text{c}}\in \mathbb{C}^{N\times L}$ is the independent fast-fading channel, $\textbf{R}_{mj}=A_j\textbf{R}_{j}\in \mathbb{C}^{L\times L}$ denotes the spatial correlation matrix of the $j$-th RIS. $A_j=d_hd_v$ is the RIS element area with $d_h$, $d_v$, the respective horizontal width and the vertical height. The $(n,m)$-th element in $\textbf{R}_j$ can be given by \cite{9598875}
 \begin{equation}
     \begin{array}{c@{\quad}c}
\displaystyle [\textbf{R}_j]_{n,m}=\text{sinc}\Bigg{(}\frac{2||\textbf{u}_n-\textbf{u}_m||}{\lambda}\Bigg{)},
     \end{array}
     \label{RIS_spatial_correlation}
   \end{equation}
where $\text{sinc}(y)=\text{sin}(\pi y)/(\pi y)$ is the sinc function, $\lambda$ is the carrier wavelength. Moreover, $\textbf{u}_x=[0,\text{mod}(x-1,L_h)d_h,\lfloor(x-1)/L_h\rfloor d_v]^T$, $x\in(m,n)$, represents the position vector\cite{9598875,10167480}. Meanwhile, the channel from the $k$-th user to the $j$-th RIS, $\textbf{g}_{kj}^{\text{c}}\in \mathbb{C}^{L\times 1}$, can be obtained by
      \begin{equation}
     \begin{array}{c@{\quad}c}
\textbf{g}_{kj}^{\text{c}}=\sqrt{\beta_{kj}^{\text{c}}}\textbf{R}_{kj}^{1/2}\textbf{v}_{kj},
     \end{array}
     \label{channel_user_RIS}
   \end{equation}
   where $\beta_{kj}^{\text{c}}$ is the relevant large-scale fading coefficient, $\textbf{v}_{kj}^{\text{c}}\in \mathbb{C}^{L\times 1}$ is the independent fast-fading channel, $\textbf{R}_{kj}=A_j\textbf{R}_{j}\in \mathbb{C}^{L\times L}$ indicates the spatial correlation of the $j$-th RIS.
   
\subsection{Electromagnetic Interference Model} 
External sources can generate the EMI as a superposition of a continuum of incoming plane waves\cite{9598875}. \cite{10167480,9598875} proved that EMI impinging on RISs has a
great effect on system performance. Electromagnetic waves are assumed to come from directions spanning large angular intervals; then, the RIS spatial correlation model can apply to isotropic scattering with uniform distribution \cite{9665300,9598875}. As such, the EMI impinging on $j$-th RIS can be modelled as
\begin{equation}
     \begin{array}{c@{\quad}c}
\displaystyle \textbf{n}_j\sim\mathcal{CN}(\textbf{0},A_j\sigma_j^2\textbf{R}_{j})
     \end{array}
     \label{EMI}
   \end{equation}
where $\sigma_j^2$ is the EMI power normalized by noise power at $j$-th RIS, and $\textbf{R}_{j}$ follows \eqref{RIS_spatial_correlation}.

\section{Uplink Channel Estimation}

During the uplink channel estimation phase, we define that $\sqrt{\tau_p}\varphi_k \in \mathbb{C}^{\tau_p\times 1}$ is the pilot sequence assigned to $k$-th user satisfying $\varphi_k^{\text{H}}\varphi_k=1,\forall k$, with $\tau_p$ representing the pilot signal length. Note that $\tau_p$ is much smaller than $\tau_c$ to improve the transmission efficiency, implying that $K>\tau_p$. Since $K>\tau_p$, some users will share the same orthogonal pilot sequences, which introduces pilot contamination \cite{5898372,9416909}. We adopt $\mathcal{P}_k$ with $\varphi_k^{\text{H}}\varphi_{k'}=1, \forall k' \in \mathcal{P}_k$, denoting the set of users using the same pilot sequence including $k$ itself.

\subsection{Two-Phase Channel Estimation Scheme} 
In order to improve the channel estimation accuracy of RIS-aided systems, \cite{9322151} proposed an ON/OFF scheme requiring high pilot overhead with $(1+JL)\tau_p$ symbols, while \cite{10225319} modified the ON/OFF scheme to reduce the pilot overhead to $(1+J)\tau_p$ symbols. Our work introduces a two-phase channel estimation scheme with fractional power control-aided pilot assignment to further balance the trade-off between the pilot overhead and the channel estimation accuracy. Our proposed scheme divides the channel estimation phase into two sub-phases requiring $2\tau_p$ symbols in total and acquires direct and cascaded RIS-aided 
channel estimates in sequence.

\subsubsection{First sub-phase} All RIS elements are OFF in the first sub-phase, and all users transmit their pilots to APs for the acquisition of the direct channel estimates between users and APs. The received pilot signal $\textbf{Y}_{m,p}^{\text{d}} \in \mathbb{C}^{N\times\tau_p}$ at the $m$-th AP is obtained by
\begin{equation}
\begin{array}{ll}
     \displaystyle \textbf{Y}_{m,p}^{\text{d}}=\sqrt{\tau_p}\sum\limits_{k=1}^{K}\sqrt{\rho_k^\text{d}}\textbf{g}_{mk}^{\text{d}}\varphi_k^{H}+\textbf{N}_{m,p}^\text{d},
\end{array}\label{received_pilot_signal_1st_phase}
   \end{equation}
where $\rho_k^\text{d}={p_k^\text{d}}/{\sigma^2}$ represents the normalized pilot signal signal-to-noise ratio (SNR) with $p_k^\text{d}$ and $\sigma^2$ representing the $k$-th user's pilot power and the noise power in the current sub-phase, respectively. $\textbf{N}_{m,p}^\text{d}$ is the additive white Gaussian noise (AWGN) at the $m$-th AP, in which the $v$-th column satisfies
$\Big{[}\textbf{N}_{m,p}^\text{d}\Big{]}_v\sim \mathcal{CN}(0,\textbf{I}_{N})$. Then, by projecting $\textbf{Y}_{m,p}^{\text{d}}$ on $\varphi_k$, we can obtain
\begin{equation}
\begin{array}{ll}
     \displaystyle \textbf{y}_{mk,p}^{\text{d}}=\sqrt{\tau_p}\sum\limits_{{k'}=1}^{K}\sqrt{\rho_{k'}^\text{d}}\textbf{g}_{m{k'}}^{\text{d}}\varphi_{k'}^{H}\varphi_k+\textbf{N}_{m,p}^\text{d}\varphi_k.
\end{array}\label{projected_pilot_signal_1st_phase}
   \end{equation}
With the assistance of the linear minimum mean-square
error (LMMSE) estimation method, the channel estimate of $\textbf{g}_{mk}^{\text{d}}$ is given by
   \begin{equation}
\begin{array}{ll}
     \displaystyle \hat{\textbf{g}}_{mk}^{\text{d}}=\frac{\mathbb{E}\Big{\{}(\textbf{g}_{mk}^{\text{d}})^H\textbf{y}_{mk,p}^{\text{d}}\Big{\}}}{\mathbb{E}\Big{\{}|| \textbf{y}_{mk,p}^{\text{d}}||^2\Big{\}}}\textbf{y}_{mk,p}^{\text{d}}=c_{mk}^{\text{d}}\textbf{y}_{mk,p}^{\text{d}},
\end{array}\label{LMMSE_direct_channel_estimation}
   \end{equation}
   where 
   \begin{equation}
\begin{array}{ll}
     \displaystyle c_{mk}^{\text{d}}=\frac{\sqrt{\tau_p\rho_k^\text{d}}\beta_{mk}^{\text{d}}}{\tau_p\sum\limits_{k'\in\mathcal{P}_k}\rho_{k'}^\text{d}\beta_{mk'}^{\text{d}}+1}.
\end{array}\label{LMMSE_direct_channel_estimation_c}
   \end{equation}

\subsubsection{Second sub-phase}In the second sub-phase, all RIS elements are ON. Distinct from the assumption adopted in \cite{9665300,10225319}, ambient electromagnetic waves in space reflected by RISs and received by APs are considered \cite{9598875,10167480}. Thus, we can express the received signal with EMI at the $m$-th AP as
\begin{equation}
\begin{array}{ll}
     \displaystyle \textbf{Y}_{m,p}=\sqrt{\tau_p}\sum\limits_{k=1}^{K}\sqrt{\rho_{k}^\text{c}}\Big{(}\textbf{g}_{mk}^{\text{d}}+\sum\limits_{j=1}^J\textbf{g}_{mkj}^{\text{c}}\Big{)}\varphi_k^{H}+\sum\limits_{j=1}^J\textbf{g}_{mj}^{\text{c}}\Theta_{j}\textbf{N}_{j}+\textbf{N}_{m,p}^{\text{c}},
\end{array}\label{received_pilot_signal_jth_phase}
   \end{equation}
where $\textbf{N}_{m,p}^{\text{c}}$ is the AWGN with $\Big{[}\textbf{N}_{m,p}^{\text{c}}\Big{]}_v\sim \mathcal{CN}(\textbf{0},\textbf{I}_{N})$, and $\rho_k^\text{c}=p_k^\text{c}/\sigma^2$ is the normalized pilot signal SNR in this sub-phase. $\textbf{N}_j$ is the EMI of $j$-th RIS, with $v$-th column satisfying
$\Big{[}\textbf{N}_j\Big{]}_v\sim \mathcal{CN}(\textbf{0},A_j\sigma_j^2\textbf{R}_{j})$. Since $\textbf{Y}_{m,p}^{\text{d}}$ is estimated in the first sub-phase and can be subtracted from $\textbf{Y}_{m,p}$, which results in
\begin{equation}
\begin{array}{ll}
     \displaystyle \textbf{Y}_{m,p}^{\text{c}}=\sqrt{\tau_p}\sum\limits_{k=1}^{K}\sum\limits_{j=1}^J\sqrt{\rho_{k}^\text{c}}\textbf{g}_{mkj}^{\text{c}}\varphi_k^{H}+\sum\limits_{j=1}^J\textbf{g}_{mj}^{\text{c}}\Theta_{j}\textbf{N}_{j}+\textbf{N}_{m,p}^{\text{c}}.
\end{array}\label{received_pilot_signal_j-th_phase}
   \end{equation}
The projection of $\textbf{Y}_{m,p}^{\text{c}}$ on $\varphi_k$ is given by
\begin{equation}
\begin{array}{ll}
     \displaystyle \textbf{y}_{mk,p}^{\text{c}}=\sqrt{\tau_p}\sum\limits_{{k'}=1}^{K}\sum\limits_{j=1}^J\sqrt{\rho_{k'}^\text{c}}\textbf{g}_{m{k'}j}^{\text{c}}\varphi_{k'}^{H}\varphi_k+\sum\limits_{j=1}^J\textbf{g}_{mj}^{\text{c}}\Theta_{j}\textbf{N}_{j}\varphi_k+\textbf{N}_{m,p}^{\text{c}}\varphi_k.
\end{array}\label{projected_pilot_signal_j-th_phase}
   \end{equation}
Given $\textbf{y}_{mk,p}^{\text{c}}$, the LMMSE of $\textbf{g}_{mk}^{\text{c}}$ is equal to
\begin{equation}
\begin{array}{ll}
     \displaystyle \hat{\textbf{g}}_{mk}^{\text{c}}=\frac{\mathbb{E}\Big{\{}(\sum\limits_{j=1}^J\textbf{g}_{mkj}^{\text{c}})^H\textbf{y}_{mk,p}^{\text{c}}\Big{\}}}{\mathbb{E}\Big{\{}|| \textbf{y}_{mk,p}^{\text{c}}||^2\Big{\}}}\textbf{y}_{mk,p}^{\text{c}}=c_{mk}^{\text{c}}\textbf{y}_{mk,p}^{\text{c}},
\end{array}\label{LMMSE_RIS_channel_estimation}
   \end{equation}
   where 
   \begin{equation}
\begin{array}{ll}
     \displaystyle c_{mk}^{\text{c}}=\frac{\sqrt{\tau_p\rho_k^c}\sum\limits_{j=1}^J\beta_{mj}^\text{c}\beta_{kj}^\text{c}\text{tr}(\textbf{T}_j)}{\tau_p\sum\limits_{k'\in\mathcal{P}_k}\sum\limits_{j=1}^J\rho_{k'}^\text{c}\beta_{mj}^\text{c}\beta_{k'j}^\text{c}\text{tr}(\textbf{T}_j)+\sum\limits_{j=1}^J\beta_{mj}^\text{c}\sigma_j^2\text{tr}(\textbf{T}_j)+1},
\end{array}\label{LMMSE_RIS_channel_estimation_c}
   \end{equation}
where
   \begin{equation}
\begin{array}{ll}
     \displaystyle \textbf{T}_j= A_j^2(\textbf{R}_{j}^{1/2})^H\Theta_j^H\textbf{R}_{j}\Theta_j\textbf{R}_{j}^{1/2}.
\end{array}
   \end{equation}
   
Then, the two-phase channel estimation-based LMMSE estimate $\hat{\textbf{g}}_{mk}$ of the aggregate uplink channel $\textbf{g}_{mk}$ can be defined as the summation of the direct channel estimates and cascaded RIS-aided channel estimates, and formulated as
\begin{equation}
\begin{array}{ll}
     \displaystyle \hat{\textbf{g}}_{mk}=\hat{\textbf{g}}_{mk}^{\text{d}}+\hat{\textbf{g}}_{mk}^{\text{c}}=c_{mk}^{\text{d}}\textbf{y}_{mk,p}^{\text{d}}+c_{mk}^{\text{c}}\textbf{y}_{mk,p}^{\text{c}}.
     \end{array}
     \label{aggregate_uplink_channel_estimation}
   \end{equation}

We adopt $\epsilon_{mk}={\textbf{g}}_{mk}-\hat{\textbf{g}}_{mk}$ as the channel estimation error
to study the channel estimation accuracy. According to \cite{10225319,8388873}, the normalized mean square error (NMSE) of the proposed scheme can be modelled as
\begin{equation}
\begin{array}{ll}
     \displaystyle\varepsilon=\frac{\sum\limits_{m,k}\mathbb{E}\Big{\{}|| \epsilon_{mk}||^2\Big{\}}}{\sum\limits_{m,k}\mathbb{E}\Big{\{}|| \textbf{g}_{mk}||^2\Big{\}}}=\displaystyle \frac{\sum\limits_{m,k}\delta_{mk}-\lambda_{mk}}{\sum\limits_{m,k}\delta_{mk}}.
     \end{array}
     \label{aggregate_uplink_channel_estimation}
   \end{equation}
with
\begin{equation}
\begin{array}{ll}
&\displaystyle\delta_{mk}=\delta_{mk}^\text{d}+\delta_{mk}^\text{c}=\beta_{mk}^\text{d}+\sum\limits_{j=1}^J\beta_{mj}^\text{c}\beta_{kj}^\text{c}\text{tr}(\textbf{T}_j),
\end{array}
\end{equation}
   \begin{equation}
\begin{array}{ll}
\lambda_{mk}
\displaystyle=\sqrt{\tau_p}\Bigg{(}\sqrt{\rho_k^\text{d}}c_{mk}^\text{d}\beta_{mk}^\text{d}+\sqrt{\rho_k^\text{c}}c_{mk}^\text{c}\sum\limits_{j=1}^J\beta_{mj}^\text{c}\beta_{kj}^\text{c}\text{tr}(\textbf{T}_j)\Bigg{)}.
\end{array}
\label{CE}
   \end{equation}  
   
\subsection{Pilot Assignment with Fractional Power Control}
\subsubsection{Fractional Power Control}We apply fractional power control depending on large-scale fading coefficients of a given user to counteract the near-far effects\cite{8968623,9322468}. Then, the pilot power of $k$-th user is
\begin{equation}
\begin{array}{ll}
     \displaystyle p_k^{\text{x}}=\frac{\sum\limits_{m=1}^M\delta_{mk}^{\text{x}}}{\sum\limits_{k'=1}^K\sum\limits_{m=1}^M\delta_{mk'}^{\text{x}}}P_{\text{sum}},
     \end{array}
   \end{equation}
where $\text{x}=\text{c},\text{d}$ represents the ongoing sub-phase, $P_{\text{sum}}=K\cdot p_p$, with $p_p$ as the pilot power without power control. 

\subsubsection{Pilot Assignment}
Similar to \cite{10264149}, we propose a fractional power control-based pilot assignment. First, for each sub-phase, the $k$-th user selects the $m_k^\text{max,x}$-th AP with the largest $\delta_{mk}^{\text{x}}$ as its prime AP,
\begin{equation}
\begin{array}{ll}
     \displaystyle m_k^\text{max,x}=\text{arg}~\text{max}_{m\in(1,...,M)}~\delta_{mk}^{\text{x}},
     \end{array}
   \end{equation}
and the $m_k^\text{max,x}$-th AP takes charge of allocating the pilot sequence $\varphi_{t_k}$ to the $k$-th user in the ongoing sub-phase. To reduce the pilot contamination to $k$-th user, for $k\leq\tau_p$, we have $t_k=k$; then, for $k>\tau_p$, we have 
\begin{equation}
\begin{array}{ll}
        \displaystyle   t_k=\text{arg}~\text{min}_{t\in\{1,2,...,\tau_p\}}~
 \sum\limits_{i\in\mathcal{S}_t}\rho_i^\text{x} \delta_{{m_k^\text{max,x}}i}^{\text{x}} ,
     \end{array}
     \label{pilot_assignment}
   \end{equation}
where $\mathcal{S}_t$ is the set of users allocated to pilot $\varphi_{t}$, and $\mathcal{S}_t$ will be updated when every $k$-th user ($k>\tau_p$) has been allocated with its pilot sequence according to \eqref{pilot_assignment}.
Moreover, $\rho_i^\text{x}=\frac{p_i^\text{x}}{\sigma^2}$ is generated by fractional power control.

\section{Downlink Data Transmission and Performance Analysis}
\subsection{Downlink Data Transmission}
After uplink channel estimation, APs transmit signals to users during the rest $\tau_d=\tau_c-2\tau_p$ symbols within each coherence interval. The downlink transmission consists of a broadcast channel making use of a precoding vector $\textbf{f}_{mk}\in\mathbb{C}^{N\times 1}$ \cite{9875036}.
In parallel, by applying the channel reciprocity of TDD operation, the uplink channel transpose can be regarded as the downlink channel\cite{9765773,8388873}. First, the transmit signal from $m$-th AP is
\begin{equation}
\begin{array}{ll}
     \displaystyle \textbf{x}_m=\sqrt{\rho_\text{d}}\sum\limits_{k=1}^K \sqrt{\eta_{mk}}\textbf{f}_{mk}q_k,
\end{array}\label{transmitted_signal_m_AP}
   \end{equation}
where $q_k\sim \mathcal{CN}(0,1)$ is the signal sent to the $k$-th user. $\rho_d=\frac{P_d}{\sigma^2}$ is the normalized SNR, where $P_d$ is the downlink transmit power at APs, and $0\leq \eta_{mk} \leq 1$ is the downlink power control coefficient selected to meet the power constraint, $\mathbb{E}\{||\textbf{x}_m||^2\}\leq \rho_d$. We consider the conjugate beamforming for downlink transmission in this work with $\textbf{f}_{mk}=\hat{\textbf{g}}_{mk}^{\text{*}}$ as the precoding vector.
Thus, the received signal at the $k$-th user considering EMI is expressed as
\begin{equation}
\begin{array}{ll}
     \displaystyle r_k&   \displaystyle =\sum\limits_{m=1}^{M}\textbf{g}_{mk}^{T}\textbf{x}_m+\sum\limits_{j=1}^{J}\textbf{g}_{kj}^{T}\Theta_j^{T}\textbf{n}_j+w_k\\&\displaystyle=\sum\limits_{m=1}^{M}\sum\limits_{k'=1}^K\textbf{g}_{mk}^{\text{T}}\sqrt{\rho_d} \hat{\textbf{g}}_{mk'}^*\sqrt{\eta_{mk'}}q_{k'}+\sum\limits_{j=1}^{J}\textbf{g}_{kj}^{T}\Theta_j^{T}\textbf{n}_j+w_{k}\\&\displaystyle=\underbrace {\sqrt{\rho_d}\sum\limits_{m=1}^{M}\sqrt{\eta_{mk}}\mathbb{E}\Big{\{}\textbf{g}_{mk}^{T}\hat{\textbf{g}}_{mk}^*\Big{\}}q_k}_\text{$\text{DS}_{\text{k}}$}\\&\displaystyle+\underbrace {\sqrt{\rho_d}\sum\limits_{m=1}^{M}\sqrt{\eta_{mk}}\Bigg{(}\textbf{g}_{mk}^{T}\hat{\textbf{g}}_{mk}^*-\mathbb{E}\Big{\{}\textbf{g}_{mk}^{T}\hat{\textbf{g}}_{mk}^*\Big{\}}\Bigg{)}q_k}_\text{$\text{BU}_{\text{k}}$}\\&+\underbrace {\sqrt{\rho_d} \sum\limits_{m=1}^{M}\sum\limits_{k'\neq k}^K\sqrt{\eta_{mk'}}\textbf{g}_{mk}^{T}\hat{\textbf{g}}_{mk'}^*q_{k'}}_\text{$\text{UI}_{\text{kk}'}$}+\underbrace {\sum\limits_{j=1}^{J}\textbf{g}_{kj}^{T}\Theta_j^{T}\textbf{n}_j}_\text{$\text{EMI}_{\text{k}}$}\\&+\underbrace {w_{k}}_\text{$\text{NS}_{\text{k}}$},
\end{array}\label{downlink_received_signal_k_user}
   \end{equation}
   where $w_k\sim \mathcal{CN}(0,1)$ is the AWGN at the $k$-th user, $\textbf{n}_j\sim\mathcal{CN}(\textbf{0},A_j\sigma_j^2\textbf{R}_{j})$ denotes the EMI of the $j$-th RIS. The terms in \eqref{downlink_received_signal_k_user} denote the respective desired signal, beamforming gain uncertainty, inter-user interference, EMI and noise.
   
\subsection{Performance Analysis}
Based on \eqref{downlink_received_signal_k_user},
the downlink SE of the RIS-aided cell-free
massive MIMO system experiencing EMI for the $k$-th user is lower bounded by \cite{10225319} 
\begin{equation}
\begin{array}{ll}
     \displaystyle \text{SE}_k=\frac{\tau_c-2\tau_p}{\tau_c}\text{log}_2\Big{(} 1+\gamma_k\Big{)},
\end{array}\label{downlink_SE}
   \end{equation}
where the effective signal-to-interference-plus-noise ratio (SINR) is $\gamma_k$ and formulated as 
\begin{equation}
\begin{array}{ll}
     \displaystyle \gamma_k \displaystyle=\frac{\mathbb{E}\{|\text{DS}_{\text{k}}|^2\}}{\sum\limits_{k'= k}^{K}\mathbb{E}\{|\text{UI}_{\text{kk}'}|^2\}-\mathbb{E}\{|\text{DS}_{\text{k}}|^2\}+\mathbb{E}\{|\text{EMI}_{\text{k}}|^2\}+1},
\end{array}\label{downlink_SE_description}
   \end{equation}
  with 
  \begin{equation}
\begin{array}{ll}
\displaystyle\mathbb{E}\{|\text{DS}_{\text{k}}|^2\}=\rho_dN^2\Big{|}\sum\limits_{m=1}^M\sqrt{\eta_{mk}}\lambda_{mk}\Big{|}^2,
\end{array}
   \end{equation}
      \begin{figure*}[t!]
    \begin{equation}
\begin{array}{ll}
     \displaystyle  \mathbb{E}\{|\text{UI}_{\text{kk}'}|^2\}=\rho_d\eta_{mk}N\Big{(}\delta_{mk}\lambda_{mk}+N\lambda_{mk}^2\Big{)}+\rho_d\sum\limits_{m=1}^M\sum\limits_{n\neq m}^M\sqrt{\eta_{mk}\eta_{nk}}N^2\lambda_{mk}\lambda_{nk}
     +\rho_d\eta_{mk'}N\delta_{mk}\Bigg{(}(c_{mk'}^\text{d})^2+(c_{mk'}^\text{c})^2\Big{(}\sum\limits_{j=1}^{J}\beta_{mj}^\text{c}\sigma_j^2\text{tr}(\textbf{T}_j)+1\Big{)}\Bigg{)}\\

 \displaystyle   +\tau_p\rho_dN\sum\limits_{m=1}^M\eta_{mk'}(c_{mk'}^\text{d})^2\Bigg{(}\underbrace{\rho_k^\text{d}(N+1)(\beta_{mk}^\text{d})^2}_{k\in\mathcal{P}_{k'}^\text{d}}+\sum\limits_{a\in\mathcal{P}_{k'}^\text{d}\setminus\{k\}}\rho_a^\text{d}\beta_{mk}^\text{d}\beta_{ma}^\text{d}+\sum\limits_{a\in\mathcal{P}_{k'}^\text{d}}\rho_a^\text{d}\beta_{ma}^\text{d}\delta_{mk}^\text{c} \Bigg{)}\\
 \displaystyle+\tau_p\rho_dN\sum\limits_{m=1}^M\eta_{mk'}(c_{mk'}^\text{c})^2\Bigg{(}\underbrace{\rho_k^\text{c}(N+1)(\delta_{mk}^\text{c})^2}_{k\in\mathcal{P}_{k'}^\text{c}}+\sum\limits_{a\in\mathcal{P}_{k'}^\text{c}\setminus\{k\}}\rho_a^\text{c}\delta_{mk}^\text{c}\delta_{ma}^\text{c}+\sum\limits_{a\in\mathcal{P}_{k'}^\text{c}}\rho_a^\text{c}\beta_{mk}^\text{d}\delta_{ma}^\text{c} \Bigg{)}\\

\displaystyle +\rho_dN^2\sum\limits_{m=1}^M\sum\limits_{n\neq m}^M\sqrt{\eta_{mk'}\eta_{nk'}}\left(\begin{aligned}&\Bigg{(}c_{mk}^\text{d}c_{mk'}^\text{d}\Big{(}\tau_p\sum\limits_{a\in\mathcal{P}_{k'}^\text{d}\cap\mathcal{P}_{k}^\text{d}}\rho_a^\text{d}\beta_{ma}^\text{d}+\underbrace{1}_{k'\in\mathcal{P}_{k}^\text{d}}\Big{)}+c_{mk}^\text{c}c_{mk'}^\text{c}\Big{(}\tau_p\sum\limits_{a\in\mathcal{P}_{k'}^\text{c}\cap\mathcal{P}_{k}^\text{c}}\rho_d^\text{c}\delta_{ma}^\text{c}+\underbrace{\sum\limits_{j=1}^J\beta_{mj}^\text{c}\sigma_j^2\text{tr}(\textbf{T}_j)+1}_{k'\in\mathcal{P}_{k}^\text{c}}\Big{)}\Bigg{)}\\
 &\times\Bigg{(}c_{nk}^\text{d}c_{nk'}^\text{d}\Big{(}\tau_p\sum\limits_{a\in\mathcal{P}_{k'}^\text{d}\cap\mathcal{P}_{k}^\text{d}}\rho_a^\text{d}\beta_{na}^\text{d}+\underbrace{1}_{k'\in\mathcal{P}_{k}^\text{d}}\Big{)}+c_{nk}^\text{c}c_{nk'}^\text{c}\Big{(}\tau_p\sum\limits_{a\in\mathcal{P}_{k'}^\text{c}\cap\mathcal{P}_{k}^\text{c}}\rho_d^\text{c}\delta_{na}^\text{c}+\underbrace{\sum\limits_{j=1}^J\beta_{nj}^\text{c}\sigma_j^2\text{tr}(\textbf{T}_j)+1}_{k'\in\mathcal{P}_{k}^\text{c}}\Big{)}\Bigg{)}\end{aligned}\right),
\end{array} 
\label{UI}
   \end{equation}
\hrulefill
\end{figure*}
     \begin{equation}
\begin{array}{ll}
\displaystyle\mathbb{E}\{|\text{EMI}_{\text{k}}|^2\}=\sum\limits_{j=1}^J\beta_{kj}^c\sigma_j^2\text{tr}(\textbf{T}_j).
\end{array}
\label{EMI_closed_form}
   \end{equation}
 The closed-form expression of the inter-user interference $\mathbb{E}\{|\text{UI}_{\text{kk}'}|^2\}$ containing the beamforming gain uncertainty can be given by \eqref{UI} at the top of the next page. According to \eqref{downlink_SE}-\eqref{EMI_closed_form}, the downlink sum SE of the proposed RIS-aided cell-free massive MIMO system with EMI can be obtained as
   \begin{equation}
\begin{array}{ll}
    \displaystyle  \text{SE}_{\text{sum}}=\sum\limits_{k=1}^K\text{SE}_k.
    \end{array}
    \label{SE_sum}
  \end{equation}

\subsection{Downlink Power Control Coefficient}
According to \eqref{transmitted_signal_m_AP}, the normalized downlink transmit power with conjugate beamforming is characterized as
\begin{equation}
 \begin{array}{l@{\quad}l}
        \displaystyle \mathbb{E}\{||\textbf{x}_m||^2\} = \rho_d \sum\limits_{k=1}^K \eta_{mk} \mathbb{E}\{ ||\hat{\textbf{g}}_{mk}||^2\}
   \displaystyle={\rho_d}N\sum\limits_{k=1}^K{\eta_{mk}}\lambda_{mk}.
     \end{array}
     \label{eq_transmitted_power_CB}
   \end{equation}

The downlink power control coefficients in this work are assumed to meet $\mathbb{E}\{||\textbf{x}_m||^2\}= \rho_d$, $\forall m$. Therefore, the potential power control coefficients for all $m=1,...,M$, $k=1,...,K$, can be given by
   \begin{equation} 
   \begin{array}{l@{\quad}l}
       \displaystyle 
\eta_{mk}=\frac{1}{KN\lambda_{mk}}.
\end{array}
\label{eq_transmitted_power_control}
   \end{equation}

      \begin{figure}[!t]
\centering
\includegraphics[width=0.83\columnwidth]{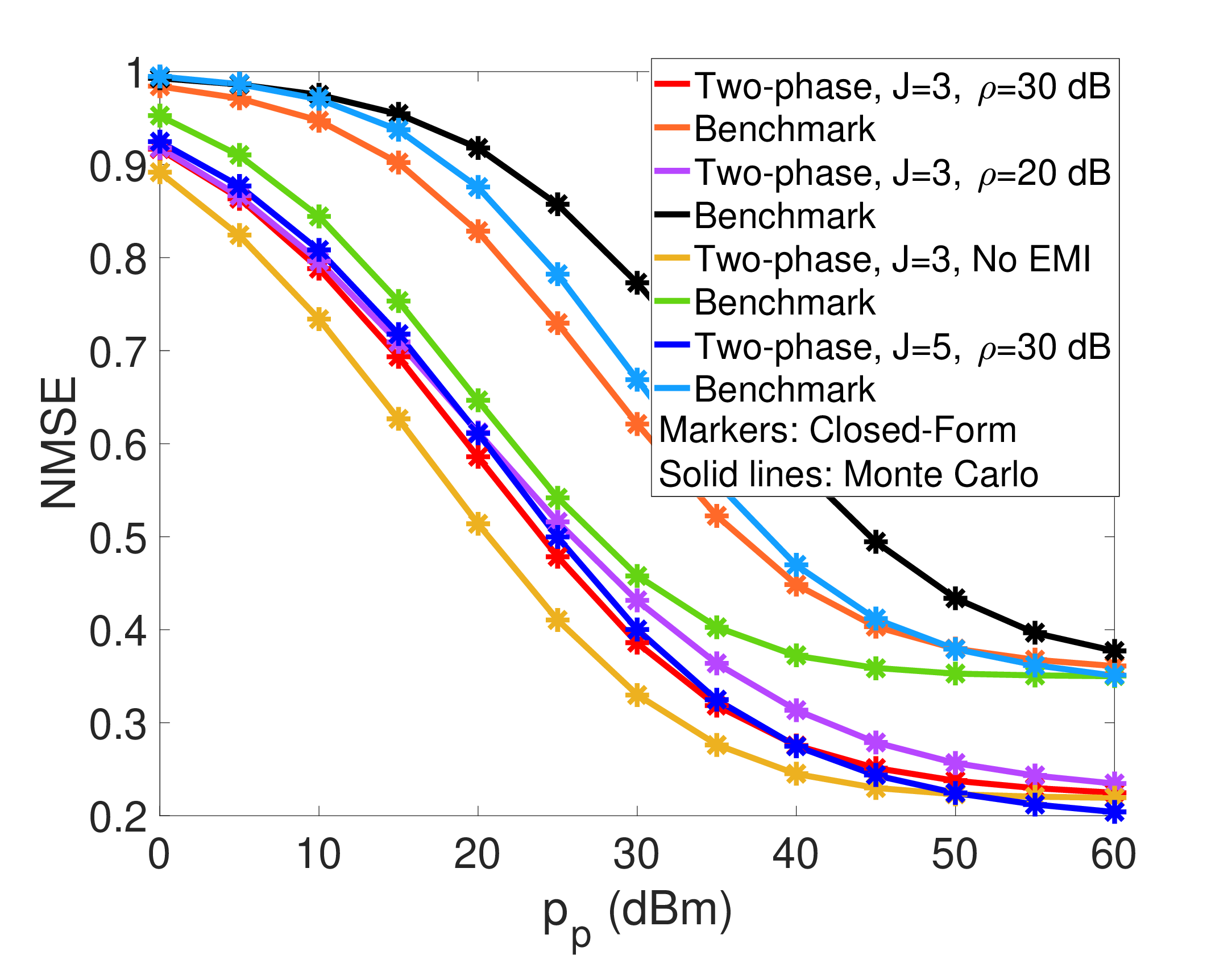}
\caption{NMSE vs. $p_p$ with $M=40$, $K=10$, $N=4$, $L=100$, $\tau_p=3$.}
\label{fig_1}
\end{figure}
 \begin{figure}[!t]
\centering
\includegraphics[width=0.83\columnwidth]{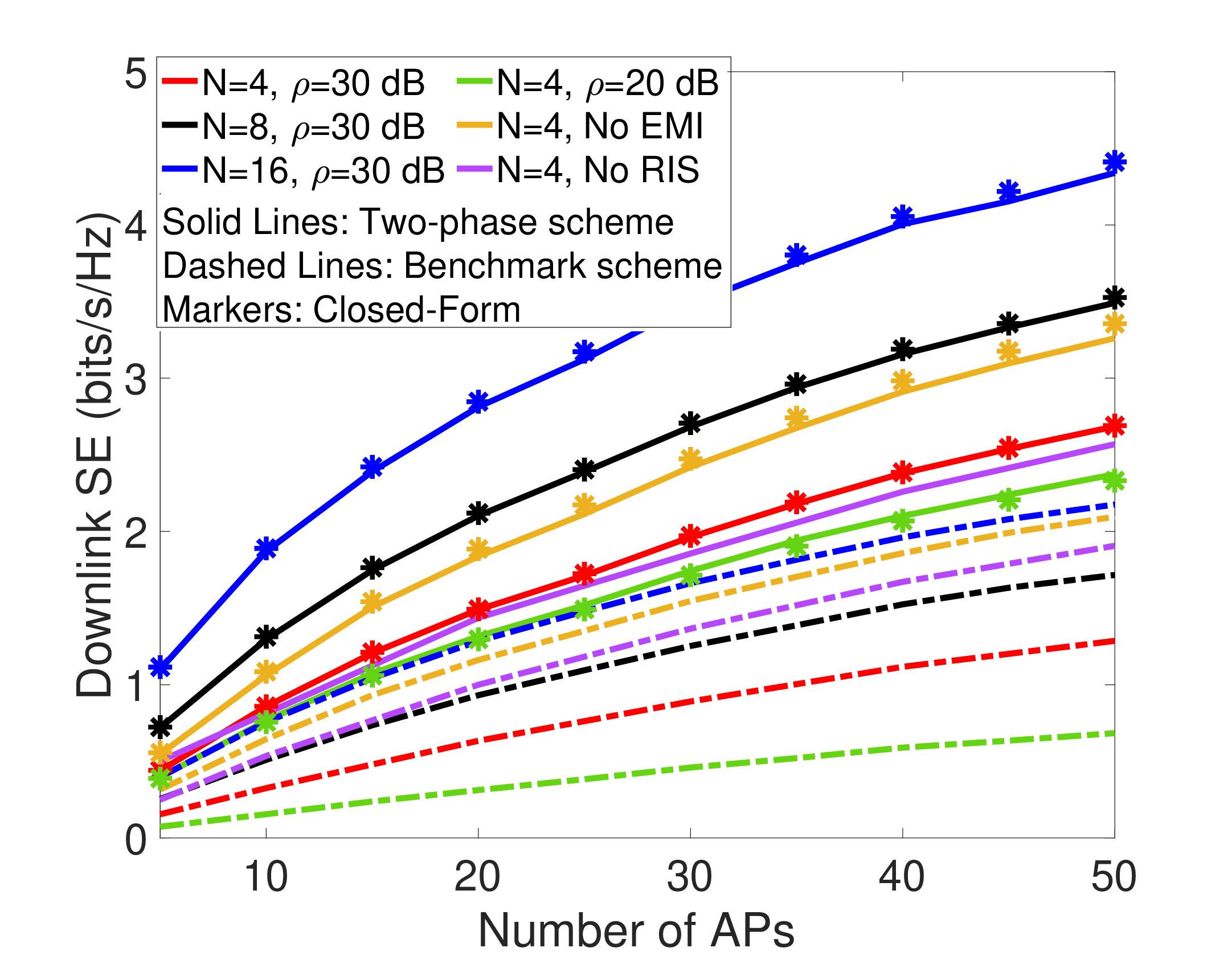}
\caption{Downlink Sum SE vs. $M$ with $K=10$, $J=3$, $L=100$, $\tau_p=3$.}
\label{fig_2}
\end{figure}
 \begin{figure}[!t]
\centering
\includegraphics[width=0.83\columnwidth]{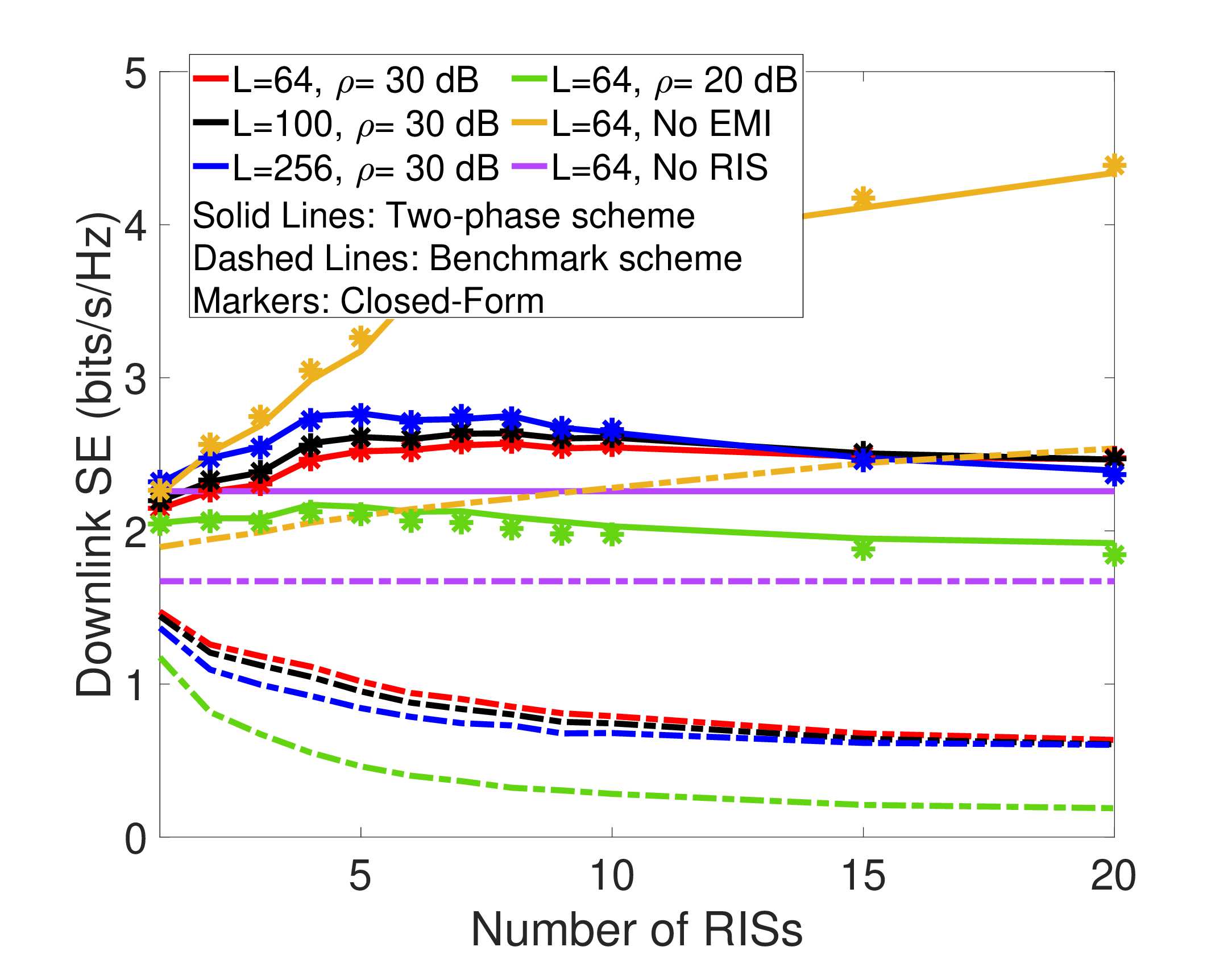}
\caption{Downlink Sum SE vs. $J$ with $M=40$, $N=4$, $K=10$, $\tau_p=3$.}
\label{fig_3}
\end{figure}
   
\section{Numerical Results}

The APs, users and RISs are assumed to be distributed within a geographic area of $D\times D$ km${^2}$, where $D=1.5$ km. Similar to \cite{9940169}, APs and users/RISs are uniformly distributed in two adjacent sub-regions such that $x^{\text{AP}},y^{\text{AP}}\in\left[-\frac{D}{2},0\right]$ km, $x^{\text{user}},y^{\text{user}}\in\left[0,\frac{D}{2}\right]$ km and $x^{\text{RIS}},y^{\text{RIS}}\in\left[0,\frac{D}{2}\right]$ km. The large-scale fading coefficients are modelled as $\beta_x=\text{PL}_x\cdot z_x$ 
($x=mk,~mj,~kj$)
, where $\text{PL}_x$ is the three-slope path loss and $z_x$ is log-normal shadowing with standard deviation $\sigma_{\text{sh}}$ [1, eq.(52)-(53)]. We set $d_0=10$
m, $d_1=50$ m, and $\sigma_{\text{sh}}=8~\text{dB} $. The AP, user and RIS heights are 15 m, 1.65 m, and 30 m \cite{10225319}.
In addition, the coherence interval contains $\tau_c=200$ symbols and $d_h=d_v=\lambda/2$ for all RISs. Until otherwise stated, we set $p_{p}=20~ \text{dBm}$, $p_{d}=23~\text{dBm}$, $\sigma^2=-91$ dBm. For EMI of $j$-th RIS, $\sigma^2_j=\displaystyle\frac{p_p\sum\nolimits_{m=1}^M\beta_{mj}^c}{\rho M\sigma^2 }$, where $\rho$ is the ratio between the received signal power and EMI power \cite{9598875,10167480}. For simplicity, we assume that all RIS elements have a fixed phase shift of $\pi/4$ and unity amplitude coefficient since weak spatial correlation introduces not significant phase shift effect of RISs\cite{10167480}. 

Fig. \ref{fig_1} shows the NMSEs of the proposed two-phase channel estimation scheme versus $p_p$. We define the commonly-used LMMSE scheme with equal power allocation as the benchmark scheme\cite{9665300,10225319}. The closed-form solutions generated by \eqref{aggregate_uplink_channel_estimation}-\eqref{CE} can closely match the simulated results. Note that $\tau_p=3$ introduces severe pilot contamination, which cannot be eliminated by increasing $p_p$, and EMI also degrades the channel estimation accuracy. Our proposed scheme can introduce additional channel estimation accuracy gain compared with the benchmark scheme to achieve a lower NMSE. However, a larger number
of RISs deteriorates the channel estimation accuracy due to the extra EMI. Moreover, the proposed channel estimation scheme might introduce extra computational complexity due to the additional estimation sub-phase; thus, the design of low-complexity channel estimation schemes in the EMI-aware
environment is left for our future work.

Fig. \ref{fig_2} studies the sum SE versus the increasing number of APs $M$ considering different numbers of antennas per AP $N$. The closed-form sum SE generated by \eqref{downlink_SE}-\eqref{SE_sum} can closely match the simulated results. Note that the two-phase scheme experiences
a smaller transmission efficiency $(\tau_c-2\tau_p)/(\tau_c-\tau_p)$ than the benchmark scheme. However, the proposed two-phase scheme can introduce extra channel estimation accuracy gain to compensate for the performance degradation and outperform the sum SE of the benchmark scheme. Moreover, increasing $M$ and $N$ can improve the SE performance. Even if EMI introduces SE degradation, the sum SE of the two-phase scheme is superior to the sum SE of the benchmark scheme without EMI and that of the RIS-free cell-free massive MIMO. These show the benefits of the proposed two-phase scheme regarding SE performance.

Fig. \ref{fig_3} shows the sum SE versus the increasing number of RISs $J$ with different numbers of elements per RIS $L$. We can find that increasing $J$ can improve the proposed two-phase scheme when $J <10$; however, the benefits of applying more RISs cannot compensate for the degradation caused by extra EMI and result in SE degradation when $J >10$. Moreover, when $J <10$, increasing $L$ can benefit the sum SE performance. In contrast, the benchmark scheme with EMI experiences an SE deterioration. Moreover, the sum SE of the two-phase scheme outperforms the sum SE of the benchmark scheme without EMI and that of the RIS-free cell-free massive MIMO under the given scenarios.


\section{Conclusion}

This work introduced a novel two-phase channel estimation scheme for the RIS-aided cell-free massive MIMO system experiencing EMI to improve the channel estimation accuracy and SE performance. We derived the closed-form downlink SE expressions to evaluate the impact of EMI and RIS spatial correlation.
Our results demonstrate that the proposed two-phase scheme could compensate for the performance degradation by introducing additional channel estimation accuracy gain in the EMI-aware environment. Moreover, increasing the number of APs could enhance the downlink SE, while the number of RISs should be selected appropriately to achieve maximum performance. These results indicate the advantages of the proposed two-phase scheme and could introduce system design guidelines to satisfy the system performance requirement considering EMI. 

\ifCLASSOPTIONcaptionsoff
  \newpage
\fi

\begin{appendices}

\end{appendices}


%




\ifCLASSOPTIONcaptionsoff
  \newpage
\fi



\bibliographystyle{IEEEtran}
\bibliography{IEEEabrv,ref}
\end{document}